\documentclass[preprint,11pt,authoryear]{elsarticle}

\usepackage{amssymb,amsfonts,graphicx,wasysym}
\usepackage{graphicx}
\usepackage{natbib}
\usepackage{amsfonts,amssymb,pifont}
\usepackage{subeqnarray}

\newcommand\p{\ensuremath{\partial}}

\newcommand\etal{\mbox{\textit{et al.}}}

\newcommand\eg{e.g.\ }

\newcommand {\ka}{\varkappa}
\newcommand {\dd}{{\rm d}}

\begin{document}

\begin{frontmatter}

\author{Igor I. Vigdorovich}
\ead{vigdorovich@imec.msu.ru}

\title{Does the flow in the viscous and logarithmic sublayers depend on the outer parameters of near-wall turbulence? How can we find the answer?}

\affiliation{organization={Institute of Mechanics, Lomonosov Moscow State University},
            addressline={Michurinsky ave. 1},
            city={Moscow},
            postcode={119192},
            country={Russia}}

\begin{abstract}
We consider turbulent flow in a half-space along an infinite plane. We regard it as the limit to which the motion in the viscous sublayer of turbulent near-wall flows tends as Reynolds number increases indefinitely. In addition to the no-slip condition, an averaged shear stress value is specified on the streamlined surface. A numerical solution algorithm is proposed, based on the well-known spectral method, which demonstrates that the boundary conditions on the wall along with the physically justified constraint that the velocity components do not grow exponentially at infinity uniquely determine the turbulent flow under consideration. The implementation of the algorithm will allow us to draw conclusions about the validity of the idea that the flow in the viscous and logarithmic sublayers is independent of the external parameters of turbulent near-wall flows.
\end{abstract}

\end{frontmatter}

\section{Introduction}
\setcounter{equation}{0} \setcounter{section}{1}

On solid surfaces in turbulent streams, due to the no-slip condition, a viscous sublayer is formed in which viscous and turbulent stresses have the same order of magnitude. We assume that a characteristic velocity in this region has the order of the friction velocity $u_\tau = \sqrt {\tau_w/\rho}$ (here, $\rho$ is the fluid density, $\tau_w$ is the averaged shear stress at the wall) while a characteristic length is calculated from the friction velocity and kinematic viscosity $\nu$. In the dimensionless variables
\begin{equation}\label{05}
x_+ = \frac {x u_\tau} \nu, ~~ y_+ = \frac {y u_\tau} \nu, ~~ z_+ = \frac {z u_\tau} \nu, ~~
t_+ = \frac {u_\tau^2 t} \nu, ~~ {\bf u}_+  = \frac {\bf u} {u_\tau}, ~~ p_+ = \frac p {\rho u_\tau^2},
\end{equation}
the flow in the viscous sublayer is described by the Navier--Stokes equations
\begin{equation}\label{001}
\frac {\p {\bf u}_+}{\p t_+} + ({\bf u}_+ \cdot \nabla)
{\bf u}_+ = - \nabla p_+ + \nabla^2 {\bf u}_+, ~~ \nabla \cdot {\bf u}_+ = 0.
\end{equation}
Since the characteristic length scale of motion in the viscous sublayer is much smaller than the external dimensions of the flow, the curvature of the streamlined surface can be neglected, considering it as an infinite plane. Then equations (\ref{001}) must be solved in the half-space $y_+ \geqslant 0$ under the no-slip conditions
\begin{equation}\label{002}
y_+ = 0: ~~ u_+ = v_+ = w_+ = 0
\end{equation}
and a given averaged shear stress on the wall
\begin{equation}\label{007}
y_+ = 0: ~~ \frac {\dd \overline u_+}{\dd y_+} = 1, ~~ \frac {\dd \overline w_+}{\dd y_+} = 0,
\end{equation}
where the overbar denotes the averaging. Second equality (\ref{007}) shows that the vector of mean turbulent motion is directed along the $x_+$ axis.

Since the extent of the region under consideration is small compared to the external scale, the mean longitudinal pressure gradient in it can be neglected (unless $\tau_w$ is very small, which is the case near a separation point, which we will not consider) and we can assume that $\overline p_+ = 0$. Then, from equations (\ref{001}) for the velocity components $u_+$ and $w_+$ and conditions (\ref{007}), we obtain
\begin{equation}\label{003}
y_+ \geqslant 0: ~~ \frac {\dd \overline u_+}{\dd y_+} - \overline {u_+ v_+} = 1, ~~ \frac {\dd \overline w_+}{\dd y_+} - \overline {v_+ w_+} = 0.
\end{equation}

Large eddies in the outer region have a significantly larger characteristic linear size. The question arises: what influence do these external parameters have on the flow in a relatively thin viscous sublayer? If this influence is exerted solely through the magnitude of the averaged shear stress $\tau_w$, then we can speak of universality of the turbulent flow characteristics in the viscous sublayer. Let $Re$ be the Reynolds number calculated from external parameters; the universality, in particular, implies the existence of a finite limit as $Re \to \infty$ with $y_+ = O(1)$ for all averaged quantities expressed in the similarity variables (\ref{05}) and the possibility of their unique determination from the solution of the problem (\ref{001})--(\ref{007}).

The universal nature of the dimensionless functions obviously extends to their asymptotic behavior as $y_+ \to \infty$. Assuming that the asymptotic behavior of the mean velocity profile depends only on two quantities~--- the distance from the wall and the friction velocity~--- Landau~\citep{Landau} derived the logarithmic law
\begin{equation}\label{04}
\overline u_+ = \frac 1 \ka \ln y_+ + C + o(1), ~~ y_+ \to \infty,
\end{equation}
in which the coefficients $\ka$ and $C$ are universal constants that do not depend on external-flow parameters.

This point of view on the behavior of the velocity profile, however, is not shared by all researchers. For example, \cite{Nagib_2008} after processing experimental data for pipe, channel and boundary layer flows concluded that the values of the von Karman constant $\ka$ and the constant $C$ depend on the flow type. And Barenblatt and his co-authors (see, \eg~\citep{Barenblatt_2014}) for many years have been developing a theory according to which (\ref{04}) should be replaced by a more complex power-law dependence with an exponent depending on $Re$. And although, as shown by \cite{Vigdorovich_2015}, Barenblatt's power-law formula cannot serve as a theoretical alternative to the logarithmic law (\ref{04}), but is simply an empirical approximation of the mean velocity profile within a limited range of Reynolds numbers, no one has yet refuted the fundamental possibility that external parameters could influence the velocity profile asymptotic behavior at the outer boundary of the viscous sublayer.

No less controversial is the question of the external parameters influence on the fluctuating motion in the viscous sublayer. The theory of "inactive" motion proposed by the renowned scientists Townsend and Bradshaw~\citep{Townsend_1961,Bradshaw_1967} explains the mechanism of this influence in the following way~\citep{Bradshaw_1994}. The motion of large eddies in the outer region generates a pressure field that induces additional fluctuating motion in the viscous sublayer, which is maintained at any Reynolds numbers. Experiments reveal a dependence of the profile of the mean-square longitudinal velocity fluctuation $u_+'^2$ on $Re$, and an increase in the maximum value of this quantity (which is reached near the wall at $y_+ \approx 15$) as the Reynolds number increases~\citep{DeGraaff_2000,Marusic_2010,Marusic_2015}. On the other hand, recently published papers by \cite{Sreenivasan_2021,Sreenivasan_2022,Sreenivasan_2023,Sreenivasan_2025} (see also \citep{Klewicki_2022}) argue that as $Re \to \infty$, this growth must cease and the quantity $u_+'^2$ will have a finite limit.

Finally, coherent structures are formed near the wall, the parameters of which (characteristic linear size, relative location, lifetime) also, according to~\cite{Smits_2011,Jimenez_2012,Massaro_2026}, depend significantly on the external parameters of the turbulent flow.

Thus, the question of what laws govern the mean velocity profile behavior and fluctuating motion in the near-wall region, whether universal, independent of Reynolds number and the type of external flow, or not, is one of the fundamental problems of turbulence theory, awaiting its solution for many decades.

Meanwhile, the answer to this question can be provided by solving the boundary value problem for the Navier--Stokes equations (\ref{001}) in the half-space $y_+ \geqslant 0$ under conditions (\ref{002}) and (\ref{007}). It does not contain any parameters and describes a flow that is statistically homogeneous with respect to $x_+$, $z_+$ and $t_+$. Next, we will give an algorithm for the numerical solution of this problem, which, in particular, shows that conditions (\ref{002}) and (\ref{007}), together with the requirement that the velocity vector components do not grow exponentially at infinity, are sufficient to uniquely determine the turbulent flow in the region under consideration. Implementation of this algorithm will provide a flow pattern in the viscous and logarithmic sublayers, based on the assumption that these regions are unaffected by external turbulent flow parameters. Comparison with experimental data and direct numerical simulation results will allow conclusions to be drawn regarding the validity of this assumption.

\section{Numerical algorithm for solving the problem}

Following \cite{Kim_1987} (see also \citep{Chevalier_2007}), we shall solve the Navier–-Stokes equations for the velocity vector components that satisfy the condition of periodicity in $x_+$ and $z_+$ with periods $L_x$ and $L_z$, respectively, by expanding the sought functions in a Fourier series in these variables. The problem in this case is reduced to integrating the equations for the expansion coefficients, which are functions of the transverse coordinate and time.

\subsection{Statement of the problem and difference equations}

Applying the divergence operator to the Navier--Stokes equations (\ref{001}) yields the Poisson equation for pressure
\begin{equation}\label{06}
\nabla^2 \left[ p + \frac 12 (u^2 + v^2 + w^2) \right] = \frac {\p H_1}{\p x} + \frac {\p H_2} {\p y} + \frac {\p H_3} {\p z},
\end{equation}
\begin{equation}\label{010}
H_1 = v \vartheta - w \omega, ~~ H_2 = w \chi - u \vartheta, ~~ H_3 = u \omega - v \chi,
\end{equation}
\begin{equation}\label{10}
\chi = \frac {\p w} {\p y} - \frac {\p v} {\p z}, ~~ \omega = \frac {\p u} {\p z} - \frac {\p w} {\p x}, ~~
\vartheta = \frac {\p v} {\p x} - \frac {\p u} {\p y}.
\end{equation}
For the sake of brevity, here and below, we omit the $+$ index at the Cartesian coordinates, time, velocity components and pressure written in the wall variables (\ref{05}). Following~\cite{Kim_1987}, we formulate the problem in terms of equations written only for the velocity vector components. The equation for the transverse vorticity component has the form
\begin{equation}\label{01}
\frac {\p \omega} {\p t} = F + \nabla^2 \omega, ~~ F = \frac {\p H_1} {\p z} - \frac {\p H_3} {\p x}.
\end{equation}
Applying the Laplace operator to equation (\ref{001}) for the transverse velocity component and taking into account equation (\ref{06}), we get
\begin{equation}\label{02}
\frac {\p \nabla^2 v}{\p t} = G + \nabla^4 v, ~~ G = \left( \frac {\p^2} {\p x^2} + \frac {\p^2} {\p z^2} \right) H_2
- \frac \p {\p y} \left( \frac {\p H_1} {\p x} + \frac {\p H_3} {\p z} \right).
\end{equation}
To obtain a closed system, we add the continuity equation to
\begin{equation}\label{9}
\frac {\p u}{\p x} + \frac {\p v}{\p y} + \frac {\p w}{\p z} = 0
\end{equation}
(\ref{01}) and (\ref{02}).

The wall boundary conditions for the system (\ref{01})--(\ref{9}) are
\begin{equation}\label{7}
y = 0: ~~ u = v = w = \omega = \frac {\p v}{\p y} = 0.
\end{equation}
They follow from the no-slip condition (\ref{002}) and continuity equation (\ref{9}). Furthermore, the physics of the problem implies that there is no exponential growth for any quantities, including velocity and vorticity
\begin{equation}\label{81}
v = o(e^{a y}), ~~ \omega = o(e^{a y}), ~~ y \to \infty, ~~ a > 0.
\end{equation}

From the asymptotic expressions for the velocity components at the wall
\begin{equation}\label{0080}
u = O(y), ~~ v = O(y^2), ~~ w = O(y), ~~ y \to 0
\end{equation}
based on (\ref{010}), we get the estimates
\begin{equation}\label{008}
H_1 = O(y^2), ~~ H_2 = O(y), ~~ H_3 = O(y^2), ~~ y \to 0.
\end{equation}

We will represent each of the periodic functions included in equations (\ref{01})--(\ref{9}) by a Fourier series segment
\begin{equation}\label{11}
h (x,\, y,\, z,\, t) = \sum_{l = -N_x/2 + 1}^{l = N_x/2 - 1} \sum_{m = -N_z/2 + 1}^{m = N_z/2 - 1}
\hat h (\alpha_l,\, y,\, \beta_m,\, t) e^{i x \alpha_l + i z \beta_m}, ~~
\alpha_l = \frac {2 \pi l}{L_x}, ~~ \beta_m = \frac {2 \pi m}{L_z}.
\end{equation}
Here, $N_x$ and $N_z$ are the number of Fourier modes used to represent the functions. Furthermore, to simplify the notation, we shall omit the indices of the wave vector components $\alpha$ and $\beta$. From (\ref{01})--(\ref{9}) we obtain the equations for the Fourier series coefficients of the functions sought
\begin{equation}\label{1}
\frac {\p \hat \omega}{\p t} = \hat F + (D^2 - k^2) \hat \omega, ~~ \hat F = i \beta \hat H_1 - i \alpha \hat H_3,
\end{equation}
\begin{equation}\label{2}
(D^2 - k^2) \frac {\p \hat v}{\p t} = \hat G + (D^2 - k^2)^2 \hat v, ~~ \hat G = - k^2 \hat H_2 - i D (\alpha \hat H_1 + \beta  \hat H_3),
\end{equation}
\begin{equation}\label{121}
i \alpha \hat u + D \hat v + i \beta \hat w = 0, ~~ D = \frac {\p}{\p y}, ~~ k^2 = \alpha^2 + \beta^2.
\end{equation}
The expressions (\ref{10}) for the vorticity vector components yield the equations
\begin{equation}\label{1200}
\hat \chi = D \hat w - i \beta \hat v, ~~ \hat \omega = i \beta \hat u - i \alpha \hat w, ~~ \hat \vartheta = i \alpha \hat v - D \hat u,
\end{equation}
resolving which, together with the continuity equation (\ref{121}), we obtain the expressions for the velocity and vorticity components in terms of $\hat \omega$ and $\hat v$, as well as of $D \hat \omega$, $D \hat v$ and $D^2 \hat v$
\begin{equation}\label{12}
\hat u = \frac i {k^2} (\alpha D \hat v - \beta \hat \omega), ~~ \hat w = \frac i {k^2} (\alpha \hat \omega + \beta D \hat v),
$$
$$
\hat \chi = \frac i {k^2} \left[ \alpha D \hat \omega + \beta (D^2 - k^2) \hat v \right], ~~
\hat \vartheta = - \frac i {k^2} \left[\alpha (D^2 - k^2) \hat v + \beta D \hat \omega \right].
\end{equation}

We will use the following finite difference approximation of equations (\ref{1}) and (\ref{2}) in time with a step $\Delta t$, as proposed by \cite{Kim_1987}:
\begin{equation}\label{42}
\left[ 1 - \frac {\Delta t} 2 (D^2 - k^2) \right] \hat \omega_{n+1} = \frac {\Delta t} 2 (3 \hat F_n - \hat F_{n-1})
+ \left[ 1 + \frac {\Delta t} 2 (D^2 - k^2) \right] \hat \omega_n,
\end{equation}
\begin{equation}\label{41}
\left[ 1 - \frac {\Delta t} 2 (D^2 - k^2) \right] (D^2 - k^2) \hat v_{n+1} = \frac {\Delta t} 2 (3 \hat G_n - \hat G_{n-1}) +
\left[ 1 + \frac {\Delta t} 2 (D^2 - k^2) \right] (D^2 - k^2) \hat v_n.
\end{equation}
The difference scheme \citep{Chevalier_2007}, which involves a more complex integration procedure with intermediate time steps, differs from (\ref{42}) and (\ref{41}) by several constant coefficients. Equalities (\ref{42}) and (\ref{41}) yield the ordinary differential equations for functions depending on the variable $y$
\begin{equation}\label{4}
(D^2 - \lambda^2) \hat \omega_{n+1} = f_n,
\end{equation}
$$
f_n = - \left( D^2 - \lambda^2 + \frac 4 {\Delta t} \right) \omega_n - 3 \hat F_n + \hat F_{n-1} = - f_{n-1} - \frac {4 \omega_n} {\Delta t}
- 3 \hat F_n + \hat F_{n-1},
$$
\begin{equation}\label{6}
(D^2 - \lambda^2) (D^2 - k^2) \hat v_{n+1} = g_n,
\end{equation}
$$
g_n = - \left( D^2 - \lambda^2 + \frac 4 {\Delta t} \right) (D^2 - k^2) \hat v_n - 3 \hat G_n + \hat G_{n-1} = - g_{n-1} - \frac 4 {\Delta t}
(D^2 - k^2) \hat v_n - 3 \hat G_n + \hat G_{n-1},
$$
$$
\lambda = \sqrt {\frac 2 {\Delta t} + k^2}.
$$

\subsection{Solution of the ordinary differential equations}

Equation (\ref{4}) is a linear, second-order, inhomogeneous equation with constant coefficients. Its general solution reads
\begin{equation}\label{8}
\hat \omega_{n+1} = C_1 e^{\lambda y} + C_2 e^{-\lambda y} + \frac {e^{\lambda y}}{2 \lambda}
\int_0^y e^{-\lambda x} f_n \dd x - \frac {e^{-\lambda y}}{2 \lambda} \int_0^y e^{\lambda x} f_n \dd x.
\end{equation}
The boundary condition (\ref{7}), according to which $\hat \omega_{n+1} = 0$ when $y = 0$, yields for the arbitrary constants the equality $C_1 + C_2 = 0$, while from the condition that $\hat \omega_{n+1}$ does not grow exponentially as $y \to \infty$, it follows that
$$
C_1 = -\frac 1 {2 \lambda} \int_0^\infty e^{-\lambda y} f_n \dd y.
$$
Now the sought function can be represented as
\begin{equation}\label{191}
\hat \omega_{n+1} = e^{-\lambda y} q [0,\, \lambda;\, f_n] - p [y,\, \lambda;\, f_n] - q [y,\, \lambda;\, f_n],
\end{equation}
\begin{equation}\label{119}
p [y,\, a;\, h] = \frac {e^{- a y}}{2 a} \int_0^y h (x) e^{a x} \dd x = \frac 1 {2 a} \int_0^y h (y - x) e^{- a x} \dd x,
\end{equation}
\begin{equation}\label{120}
q [y,\, a;\, h] = \frac {e^{a y}}{2 a} \int_y^\infty h (x) e^{- a x} \dd x = \frac 1 {2 a} \int_0^\infty h (y + x) e^{- a x} \dd x.
\end{equation}
Here, $h$ is generally a sign-changing function growing at infinity more slowly than exponentially. Therefore, the improper integral (\ref{120}) converges for $a > 0$. As follows from (\ref{119}) and (\ref{120}), the linear functionals $p$ and $q$ satisfy the first-order equations
\begin{equation}\label{19}
D p = - a p + \frac h {2 a}, ~~ D q = a q - \frac h {2 a},
\end{equation}
using which after differentiating (\ref{191}), we obtain
\begin{equation}\label{192}
D \hat \omega_{n+1} = \lambda \left\{ p [y,\, \lambda;\, f_n] - q [y,\, \lambda;\, f_n] - e^{-\lambda y} q [0,\, \lambda;\, f_n] \right\}.
\end{equation}

For $k > 0$, the general solution of the fourth-order equation (\ref{6}) is given by
\begin{equation}\label{129}
\hat v_{n+1} = C_1 e^{\lambda y} + C_2 e^{-\lambda y} +  C_3 e^{k y} + C_4 e^{-k y}
$$
$$
+ (\lambda^2 - k^2)^{-1} \{ p [y,\, k;\, g_n] + q [y,\, k;\, g_n] - p [y,\, \lambda;\, g_n] - q [y,\, \lambda;\, g_n] \},
\end{equation}
which, using (\ref{19}), can be verified by direct substitution. Since the functions $p [y,\, k;\, g_n]$, $q [y,\, k;\, g_n]$, $p [y,\, \lambda;\, g_n]$ and $q [y,\, \lambda;\, g_n]$ grow more slowly than exponentially, $C_1 = C_3 = 0$. Conditions (\ref{7}), according to which $\hat v_{n+1} = 0$ and $D \hat v_{n+1} = 0$ at $y = 0$, give the equations for the remaining two arbitrary constants
$$
C_2 + C_4 = \frac {q [0,\, \lambda;\, g_n] - q [0,\, k;\, g_n]} {\lambda^2 - k^2}, ~~
\lambda C_2 + k C_4 = \frac {k q [0,\, k;\, g_n] - \lambda q [0,\, \lambda;\, g_n]}{\lambda^2 - k^2},
$$
after solving which the expression (\ref{129}) can be represented as
\begin{equation}\label{13}
\hat v_{n+1} = \frac {\Phi [y,\, k,\, \lambda;\, g_n] - \Phi [y,\, \lambda,\, k;\,\, g_n]}{\lambda^2 - k^2},
\end{equation}
\begin{equation}\label{130}
\Phi [y,\, a,\, b;\, g_n] = p [y,\, a;\, g_n] + q [y,\, a;\, g_n] + \{(a + b) q [0,\, a;\, g_n] - 2 b q [0,\, b;\, g_n] \}
\frac {e^{-a y}}{a - b}.
\end{equation}

To compute the quantities (\ref{12}), we need the first and second derivatives of $v_{n+1}$, after calculating which from (\ref{13}) and (\ref{130}) using (\ref{19}), we have
\begin{equation}\label{131}
D \Phi [y,\, a,\, b;\, g_n] = a q [y,\, a;\, g_n] - a p [y,\, a;\, g_n] + \{2 b q [0,\, b;\, g_n] - (a + b) q [0,\, a;\, g_n]\}
\frac {a e^{-a y}}{a - b},
\end{equation}
$$
(D^2 - k^2) \hat v_{n+1} = - \Phi [y,\, \lambda,\, k;\, g_n].
$$

The quantity $\hat G$ in (\ref{2}) contains the derivatives $D \hat H_1$ and $D \hat H_3$. Since according to (\ref{008}), $\hat H_1$ and $\hat H_3$ vanish at the wall, from the representations (\ref{119}), \eg for $\hat H_1$ after integration by parts, we obtain
\begin{equation}\label{009}
p [y,\, a;\, D \hat H_1] = \frac {\hat H_1}{2 a} - a p [y,\, a;\, \hat H_1], ~~ q [y,\, a;\, D \hat H_1] =
- \frac {\hat H_1}{2 a} + a q [y,\, a;\, \hat H_1].
\end{equation}
We also need an expression for the second derivative $D^2 \hat v_n$, which forms part of the quantity $g_n$. Applying relations (\ref{009}) twice, taking into account the wall conditions $\hat v_n = D \hat v_n = 0$, we have
$$
p [y,\, a;\, D^2 \hat v_n] = \frac {D \hat v_n}{2 a} - \frac {\hat v_n} 2 + a^2 p [y,\, a;\, \hat v_n], ~~
q [y,\, a;\, D^2 \hat v_n] = - \frac {D \hat v_n}{2 a} - \frac {\hat v_n} 2 + a^2 q [y,\, a;\, \hat v_n].
$$
These relations along with (\ref{130}) and (\ref{131}) give
$$
\Phi [y,\, a,\, b;\, D^2 \hat v_n] = - \hat v_n + a^2 \{p [y,\, a;\, \hat v_n] + q [y,\, a;\, \hat v_n]\}
$$
$$
+ \{(a + b) q [0,\, a;\, \hat v_n] - 2 b q [0,\, b;\, \hat v_n] \} \frac {a^2 e^{-a y}}{a - b},
$$
$$
D \Phi [y,\, a,\, b;\, D^2 \hat v_n] = - D \hat v_n + a^3 \{q [y,\, a;\, \hat v_n] - p [y,\, a;\, \hat v_n]\}
$$
$$
+ \{2 b q [0,\, b;\, \hat v_n] - (a + b) q [0,\, a;\, \hat v_n]\} \frac {a^3 e^{-a y}}{a - b},
$$
which implies the following equalities:
$$
(\lambda^2 - k^2)^{-1} \{\Phi [y,\, k,\, \lambda;\, (D^2 - k^2) \hat v_n] - \Phi (y,\, \lambda,\, k;\,\, (D^2 - k^2) \hat v_n]\} =
- \Phi [y,\, \lambda,\, k;\, \hat v_n],
$$
$$
(\lambda^2 - k^2)^{-1} D \{\Phi [y,\, k,\, \lambda;\, (D^2 - k^2) \hat v_n] - \Phi (y,\, \lambda,\, k;\,\, (D^2 - k^2) \hat v_n]\} =
- \lambda \Phi [y,\, \lambda,\, k;\, \hat v_n].
$$

\subsection{Zero wave number}

Conditions (\ref{007}) and (\ref{003}), which specify the fluid motion, are formulated for averaged quantities that are calculated at $\alpha = \beta = 0$, when the above representations of the sought functions are unsuitable. In this case, the formula for the Fourier series coefficient, representing, \eg the longitudinal velocity component, reads
\begin{equation}\label{40}
\hat u_0 (y,\, t) = \frac 1 {L_x L_z} \int_0^{L_x} \int_0^{L_z} u (x,\, y,\, z,\, t) \dd x \dd z.
\end{equation}
Here, the index $0$ indicates that the quantity is calculated at zero wave number.

The algorithm for solving the problem consists of sequentially integrating the non-stationary Navier--Stokes equations until a flow that is statistically homogeneous in time is achieved. We shall use relations (\ref{003}) at each time step, calculating the averaged quantities using formulas analogous to (\ref{40}) as integrals over the spatial variables $x$ and $z$. Then, relations (\ref{003}) can be written as
\begin{equation}\label{21}
D \hat u_0 (y,\, t) - \widehat {u v}_0 (y,\, t) = 1, ~~ D \hat w_0 (y,\, t) - \widehat {v w}_0 (y,\, t) = 0.
\end{equation}
When $\alpha = \beta = 0$, the continuity equation (\ref{2}) takes the form $D \hat v_0 = 0$, which, given the wall boundary condition, yields $\hat v_0 \equiv 0$. This means that to calculate the integrals of the velocity components products $\widehat {u v}_0$ and $\widehat {v w}_0$, using the series (\ref{11}), one does not need to know the quantities $\hat u_0$ and $\hat w_0$, which define the mean velocity profiles. These profiles can be calculated from (\ref{21})
$$
\hat u_0 (y,\, t) = \int_0^y [1 + \widehat {u v}_0 (y,\, t)] \dd y, ~~ \hat w_0 (y,\, t) = \int_0^y \widehat {v w}_0 (y,\, t) \dd y.
$$
Moreover, from (\ref{1200}) we obtain
$$
\hat \chi_0 = D \hat w_0, ~~ \hat \omega_0 \equiv 0, ~~ \hat \vartheta_0 = - D \hat u_0 (y,\, t) = - 1 - \widehat {u v}_0 (y,\, t).
$$

\section{Concluding remarks}

Let us note the qualitative features of the flow under consideration. It differs from channel flows, since it occupies an unbounded region~--- a half-space~--- and from boundary-layer flows, since, unlike the latter, it has an unlimited velocity at infinity. These features are reflected in the formulation of the boundary conditions, which are specified only on the streamlined surface in the form of an averaged shear stress and zero velocity. Together with the requirement of the absence of the velocity components exponential growth at infinity, these conditions prove sufficient to uniquely determine the turbulent flow.

Solving the problem at each time step is reduced to the calculation of two functions of the transverse coordinate $p [y,\, a;\, h]$ and $q [y,\, a;\, h]$, defined by the integrals (\ref{119}) and (\ref{120}). This operation is most expediently performed by integrating the first-order ordinary differential equations (\ref{19}) that these functions satisfy. In this case, the initial condition at the wall
$$
q [0,\, a;\, h] = \frac 1 {2 a} \int_0^\infty h (x) e^{- a x} \dd x = \frac 1 {2 a^2} \int_0^\infty h \left( \frac x a \right) e^{- x} \dd x
$$
can be determined using the well-known technique of calculating improper integrals with the weight function $e^{- x}$ based on the values of the integrand at the roots of the Laguerre polynomial~\citep{Krylov_1967}.

\end{document}